\documentclass[fonts]{icst}

\usepackage{moreverb}

\usepackage[breaklinks=true,colorlinks,bookmarksopen,bookmarksnumbered,linkcolor=ICSTblue,citecolor=blue,urlcolor=ICSTblue]{hyperref}
\usepackage{breakurl}
\usepackage{doi}

\newcommand\BibTeX{{\rmfamily B\kern-.05em \textsc{i\kern-.025em b}\kern-.08em
T\kern-.1667em\lower.7ex\hbox{E}\kern-.125emX}}

\def\volumeyear{2014}

\journalname{XXXXXX}
\articletype{Research Article}
 \def\volumeyear{XXXX}
 \def\volumenumber{XX}
  \def\issuemonth{XX-XX}
  \def\issuenumber{X}
  \def\articleid{XX}
  \def\copyrightholder{Nayak\textit{ et al.}, licensed to ICST}
  \copyrightnote{This is an open access article distributed under the terms of the Creative Commons Attribution license (\url{http://creativecommons.org/licenses/by/3.0/}), which permits unlimited use, distribution and reproduction in any medium so long as the original work is properly cited.}
  \def\articledoi{10.4108/XX.X.X.XX}
\received{XXXX}
  \accepted{XXXX}
  \published{XXXX}

\begin{document}

\runningheads{Nayak et. al.}{Big Computing: Where are we heading?}

\title{Big Computing: Where are we heading?}

\author{Sabuzima Nayak\affil{1}, Ripon Patgiri\affil{1}\fnoteref{1} and Thoudam Doren Singh\affil{1}}

\address{\affilnum{1}National Institute of Technology Silchar, Assam-788010, India}

\abstract{This paper presents the overview of the current trends of Big data against the computing scenario from different aspects. Some of the important aspect includes the Exascale, the computing power and the kind of applications which offer the Big data. This starts with the current computing hardware constraint against the need of the rising Big data applications. We highlight the issues and challenges of energy requirement, software complexity, hardware failure, fault tolerant computing, and communication. As the complexity of computation is going to rise in the future. The paper also highlights the future direction of Big computing systems for Bioinformatics, social media, hardware and software requirements, data intensive computation and then towards GPU era.}

\keywords{Big Computation, Big Computing, Big Data, Exascale, Exaflops, Zettaflops, GPU.}


\fnotetext[1]{Corresponding author.  Email: \email{ripon@cse.nits.ac.in}}

\maketitle

\section{Introduction}
Big Computing is the future trend in two-fold, namely, data-intensive computation (Big Data) and task-intensive computation. Big Data deals with data-intensive computation while High Performance Computing (HPC) deals with task-intensive computation. If the scientific world is unable to produce a Exascale supercomputer (HPC) in the next few years, then the avalanche of data produced at an Exascale speed cannot be processed. Currently, the world is producing data of nearly 2.5 Exabytes per day as reported by the IBM \cite{IBM}. Hence, the size of Big Data in the near future is going to reach Exascale. The current Big data technologies are unable to handle such Exascale data. If the world is not prepared for the future, then it will lead to shutting of data processing and loss of information. Storage of data is not the aim of the technology, but to process the stored data and utilize the mined knowledge. Hence, the Exascale Big data require a Big computing system capable of processing Exabyte sized data. 

\begin{figure*}
    \centering
    \includegraphics[width=0.8\textwidth]{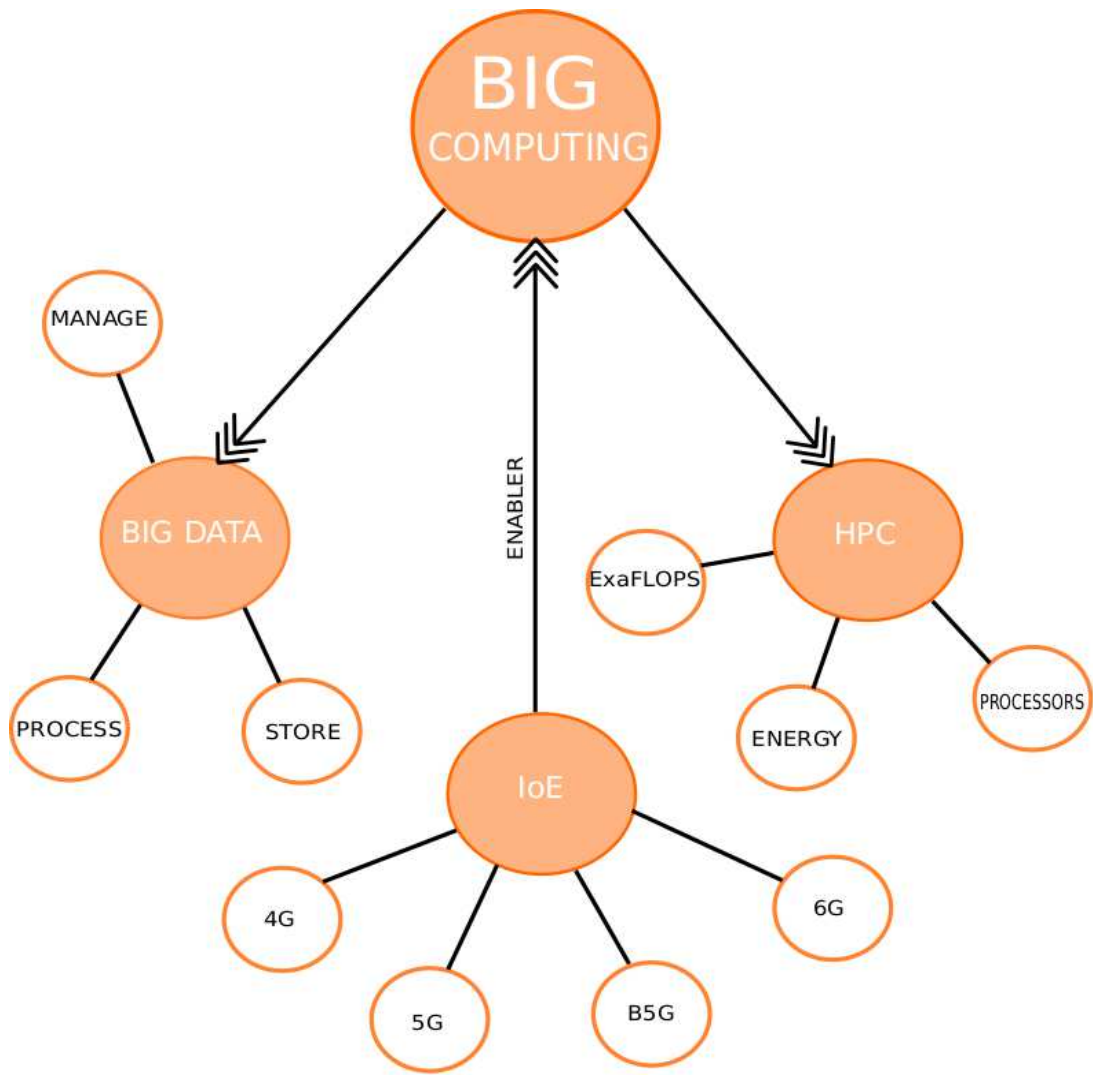}
    \caption{Big Computing and its enabling technologies}
    \label{fig1}
\end{figure*}

The future is about Exascale which is depicted in Figure \ref{fig1}. Hence, no super power country wants to be left behind in the race of producing Exascale computational system. Europe has launched EuroHPC initiative to develop a supercomputer based on European technology by 2023 and to become an independent superpower in the race of Exascale. It has provided a funding of 1 billion euro till 2020 \cite{Gagliardi}. Similarly, the United States is investing \$1.8 billion USD for development of three Exascale Big computing systems till 2021-2022 \cite{Timothy}. In TOP500, United states Summit \cite{summit} supercomputer achieved the first position on June 2018. Summit has 9216 CPU (Central processing unit) chips and tightly connected 27648 GPUs (Graphics processing Unit). TOP500 \cite{top500} is an event on advanced digital technology that ranks top 500 powerful computers. For ranking the computer systems, LINPACK Benchmark \cite{linpack} is followed which measure the ability to solve dense linear equations problem set using floating point arithmetic. For every country having its own supercomputer is not just a competition, but answers to many unsolved and complex problems. Summit is used to study about human diseases, fusion energy, advanced materials and many more areas \cite{PALMER}.

The article is organized as follows- Section \ref{bd} briefs on Big Data systems. Section \ref{hpc} explores on High Performance Computer. Section \ref{ic} presents rich insight on issues and challenges to be faced by Big computing system in future. Section \ref{zf} exposes future possibilities of ZettaFLOPS systems. Section \ref{fd} elaborates future aspects of Big computing systems. Finally, the article draws a suitable conclusion in Section \ref{con}.

\section{Big Data}
\label{bd}
The current trend of computing scenario has introduced data explosion in different facets of life. Thus, Big Data is one of the most popular game changer paradigm. It has no untouched area, particularly, government sector, public sector, and private sector \cite{RA}. More specifically, Big Data is applied in business, economy, healthcare, medicine, bioinformatics, earthquake, weather, stock market, and many emerging areas. The growing revenue is amazing in Big Data technology. Therefore, billion dollars are invested in Big Data technology. Moreover, the data are to grow at an exponential pace. In 2009, the Big Data started to grow at an exponential pace till 2019. However, the data growth rate will not be exponential after 2020. The data growth rate will face declination from 2025 to 2030. The data growth rate will be at a linear pace rather than exponential pace. However, the growth rate of data is expected to grow constantly at a linear pace in future. There is a requirement for Big computing system to compute massive scale dataset. Big Computing should redefine Big Data such that every company can be able to compute their Exabytes of dataset at their own farmhouse in the near future. Almost, all IT industry has to face Exabytes of data after 2025. Hence, Big Data 2.0 will be face off in Big Computation. Big Data 2.0 is yet be conceived. Thus, Industry will completely transform from Big Data to Big Data 2.0 completely after 2025. The Big Data 2.0 has to deal with massive amount which includes exabytes and beyond. Current state-of-the-art unable to deal with massive amount of data which scale exabytes or beyond. It is expected that Big Data 2.0 will be evolved in 2025. As data are growing at an exponential pace, the Internet of Everything (IoE) will generate enormous amount of data as shown in Figure \ref{fig1}. From 2020 to 2025, Beyong 5G (B5G) will evolve and hence, the Big Data 2.0 is required to for processing massive amount of data.

\section{High Performance Computing}
\label{hpc}
Essentially, the knowledge or information extraction from the Big data lies in the way how the high performance computation is carried out. ZettaFLOPS is the next era for HPC which can perform  $10^{21}$ floating point operations per second (FLOPS). Current race is with ExaFLOPS in HPC which can perform $10^{18}$ FLOPS. Many countries are participating to get a position in TOP500. ZettaFLOPS is yet to be achieved. It is expected to achieve by around 2025. Hence, the scientific and research world is focusing on creating a supercomputer/High-Performance Computing (HPC) which can perform beyond ExaFLOPS. However, there are many issues that need to be dealt before creating a ZettaFLOPS system.

\section{Issues and challenges}
\label{ic}
\subsection{Energy requirement}
A Big computing system approximately requires processing capability of 50 GigaFLOPS/Watt of power. Currently, we are unable to produce energy efficient computation. Energy is consumed by processors, cooling fans, main board, network infrastructure and many other components. Moreover, the voltage supply is reduced to leverage the energy consumption, but it results in data corruption \cite{Martino}. Power consumed by datacenter, and storage infrastructure also add to the total power consumption. Farmhouse bears more cost for energy compared to establishing entire farmhouse. In other scenario, Data are oil of the modern epoch. Therefore, data are being amalgamated which forms a data silo. This data silo requires huge processing power to mine meaningful insight. Data is growing and therefore, hardware is augmented to the datacenter to increase scalability. Thus, datacenter requires more power supplies. Therefore, energy-aware computation is the next emerging area.

\subsection{Hardware failure}
Many node failure occurs due to hardware failure. Currently, suppose a single processor of a system takes 25 years for mean time to failure. Then, in a Big computing system with hundred thousands of the processor has a mean time to failure of 2 hours \cite{Bautista}. The throughput of each component is increased by using transistor scaling process. Transistor scaling is embedding more transistors in each computing device. But decreasing feature size leads to reduction in critical charge. Critical charge is used to change the logic of a circuit. So, the change results in data corruption. It is called soft errors in Big computing systems. In addition, DRAM (dynamic RAM) errors lead to undetected errors. Those errors results in incorrect answer calculation. 

\subsection{Fault tolerance}
Big computing systems have millions of components with frequent failures. In such cases, the network should provide efficient fault-tolerant mechanisms. In software layer, the upper level layer should provide fault-tolerance mechanisms to make robust software infrastructure. To devise such mechanism, deep learning and streaming will play a big role.

\subsection{Communication}
Communication among the component requires energy. Hence, reduction in the number of communication reduces power consumption. But, in the case of Big computing system, the communication is between hundreds thousands of components. Hence, reducing the number of communication is not an easy task. Currently, zero-round trip time (0-RTT) is emerging \cite{0RTT}. However, in case of occurrence of failure, it affects the communication among the components. So fault-tolerant techniques among the components are the necessity. From another viewpoint, network communication among millions of nodes in an Exascale distributed system need to be efficient and fault-tolerant. Moreover, reduction of latency is also dependant of efficient network communication. Besides, 5G is emerging. 5G will be able to provide 20 GBPS downlink peak data rate, and 10 GBPS uplink peak data rate. Furthermore, 5G will revolutionize almost all major industries. Thus, Big Computing will become a more prominent player in modern IT industries.

\subsection{Software Complexity}
In Big computing system, the complexity of system and application software increases to many folds \cite{geist}. First, increasing computational efficiency. New complex software is required to include finer temporal and spatial scales. This needs heavy data assimilation and very complex physics. Second, adding algorithmic resilience. It eliminates undetected errors. Third, energy-awareness. It is the most important necessity because very high energy consumption is making the existence of the Big computing system impossible. Fourth, distributed system. Storing Exascale data is not possible by using a single datacenter. Therefore, the programming model is required to re-engineered for the multicore CPU (e.g. data parallel MPI (Massage Passing Interface) \cite{mpi}) and another for the accelerator (e.g. OpenCL \cite{opencl}). Fifth, memory levels. The hierarchy of memories increases the software complexity.

\section{ZettaFLOPS}
\label{zf}
Achieving Exascale, Big computing system requires efficient hardware to handle millions of computing per second. Performance of a Big computing system is evaluated by the number of floating point operations per second (FLOPS). In this section, some new approaches are discussed from the hardware aspect. 

\subsection{Processor}
The processing unit is the main heart of the Big computing system. To design a Big computing system, a high-performing processor with efficient architectural design is essential. Currently, high-performing processor are implementing multicore and SIMD (single instruction multiple data) techniques to reach the performance target. Heterogeneous design is also considered. It combines a general-purpose multicore processor and  a many-core coprocessor. Multicore processor performs irregular computation and many-core coprocessor performs high throughput computation. Till date, the Exascale performing processor is not achieved, but stitching the new processor technology \cite{shalf} will make it possible to achieve the Exascale standards. However, many issues and challenges has to be resolved to achieve a Big Computing processor. Some issues are (a) Exascale level performance, (b) power efficient, (c) satisfying the need of new applications such as deep learning and Big data analytics. Hence, it is important to realize that a processor has to balance performance and energy for Big computing system. 

ARM \cite{ARM} processors are currently gaining popularity because power consumption is one of the main issues in designing the Big computing systems. Advanced RISC Machines (ARM) are embedded in many smart phones and tablets. ARM \cite{ARM} uses Restricted Instruction Set Computer (RISC) instructions. RISC instruction sizes are fixed, has simple instruction fetch, and simultaneous access to the opcode and operand. This makes the control unit design simple and consumes less power. Moreover, RISC instructions execute in one clock cycle, which reduces the interrupt latency. Cost of server-oriented ARM SoC (single on-chip) is less compared to Intel x86 CPUs. In addition, ARM is easy to debug. However, ARM has some issues \cite{lobet} such as efficiency of ARM reduces in case of strong scalable vector usage, and the architectural design is complex. On the other hand, it is estimated that ARM is capable of replacing x86 and POWER-based servers with some optimization in Big computing systems and hyperscale data centers \cite{EUROSERVER}. 

ExaNest \cite{exanest} is a computing architecture designed for Exascale-class systems. It is built upon 64 bit ARM processors. ARM processors \cite{ARM} utilize 2-3 times less electricity and performs more computation. ExaNest reduces power consumption by decreasing:- data transfer distance, footprint area and cost of installation, and the number of devices required to reach performance ambition. The storage devices are kept close to compute nodes and connected using fast custom-made interconnects. A single and unified interconnect is implemented to manage storage and application traffic. A daughter board contains the basic compute unit to enhance the performance of computation. It has 4 Xilinx Ultrascale+ FPGAs, 16 cores (each Xilinx Ultrascale+ FPGA with quad ARM Cortex-A53 64-bit), every FPGA is attached to 16 gigabytes of DDR4 memory, and an NVM in-node SSD for storage.

\subsection{Storage memory}
Many new memory technologies are proposed to replace DRAM. These technologies should be able to solve many problems such as less power consumption, achieving higher parallelism, run-time error handling and many more. In addition, Big computing system will have a heterogeneous memory and storage model. For example, along with DRAM non-volatile memory (NVM) can be used to increase the efficiency of I/O operations \cite{panda}. Similarly, Phase change memory (PCM) \cite{PCM} uses phase-change material to store data. The phase-change material is either amorphous or crystalline. It provides superior density compared to DRAM. Moreover, the PCM can have different degree of crystallization in different cells. This enables each cell to store more than one bit \cite{bedeschi}. PCM also scale better compared to DRAM \cite{freitas}. But it has some demerits such as it is slower compared to DRAM, has more memory latency, and performs less write operations to reduce latency. Another technology is STT-MRAM (Spin-Transfer Torque Magnetic Random Access Memory) array \cite{SST} based on Magnetic Tunneling Junction (MTJ). An MTJ is a thin tunneling having a dielectric in between two ferro-magnetic layers. STT-MRAM are compatible with the conventional DRAM chips and can be used to construct byte-addressable memory devices. It is capable of replacing DRAM. However, STT-MRAM require quick reads which causes problems due to small sense margins. During manufacturing, thermal fluctuations causes high write errors. In Big computing system,  efficiency in the main memory is mainly required. Hence, an upgraded and highly efficient main memory will be able to handle and manage millions of operations per seconds.  

SAGE (percipient StorAGe for Exascale data centric computing) \cite{NARASIMHAMURTHY} is an Exascale capable hardware. It has a multi-tiered I/O hierarchy with an intelligent management software. SAGE has a software stack based on Unified Object-Based Storage Infrastructure. The software stack consists of mainly three layers, namely, Mero, Clovis and third layer consist of tools and HPC APIs. Mero is the lowest layer in the software stack. It provides a platform for distributed object storage. Mero comprise of a core that supports basic object storage features. It also does resource management (locks, caches, extents, etc.). In addition, Mero provides hardware reliability, for example, RAID enabled through Server Network Striping. Clovis is the second layer which is a transactional storage API. User applications directly use this layer. In the third layer, SAGE uses some tools for I/O profiling and optimizing data movement. Apache Flink tools enable data analytic jobs. SAGE takes the support of POSIX compliant storage access. SAGE uses HSM to control data movement in the SAGE hierarchies using the data usage information. Moreover, SAGE has developed a new tool called RTHMS \cite{Peng}. RTHMS analyses parallel applications to recommend about data placement of memory objects in the heterogeneous memory systems. It is designed to handle Exascale operations, however, many new features can be added to increase the efficiency.

\section{Future direction}
\label{fd}
The Big computing systems cannot reach its expected performance level by using the existing methods and techniques. Making the Big computing system a reality, new or hybrid approaches are required. In this section, new approaches that are proposed for Exascale system is discussed. 

\subsection{Communications}
As depicted in Figure \ref{fig1}, the enabler of Big Computing is Internet of Everything (IoE). The IoE is backed by mobile communication. Currently, 5G mobile communication is being deployed in many countries, and also, Beyond 5G (B5G) is being developed. Moreover, it is expected that 6G mobile communication will be deployed from 2030 and onward. The 6G will revolutionize modern lifestyle, society, business and industry, and thus, we will evidence the transition from Internet of Things (IoT) to IoE \cite{Nayak}. Moreover, there will be many transitions, for instance, smart devices to intelligent devices. Also, 6G will be able to deliver many new technologies, for instance, holographic communication, five sense communication, intelligent vehicles, etc., to provide rich Quality of Experiences (QoE) \cite{Nayak}. Due to 6G technology, IoE will create a massive small-sized data. These massive small-sized data will pose a challenge to Big Computing in storing, processing and managing. Therefore, Big Computing is to be redefined for massive scale small-sized data. 

\subsection{Bioinfromatics}
With the advancement of DNA sequence extraction technology (next-generation sequencing) and Big Data \cite{stephens} resulted in decreasing the cost of whole genome sequencing to a reasonable price \cite{bioExascale}. However, the data obtained are unordered small fragments of the sequence numbered to billions. These data scale to terabytes. However, with the help of next-generation sequencing, genomes of all non-human organism will be obtained in the future and the memory requirement is in Exascale or beyond. The DNA collecting repository currently has a size of petabyte \cite{stephens} such as The Cancer Genome Atlas (TCGA) and Human Microbiome Project. The size will increase and reach Exascale in near future.

\subsection{Social Media}
Today, social media is another area that uses Big Data technology for storage, processing and analyzing data. The user stores their messages, photos, music, video, etc. in the storage provided by the social media service provider. These data are mostly public so it opened the door for the businesses to analyze these data and understand their customers. These data provide the demography, cultural preferences, likes, dislikes, etc. These data are analyzed to obtain the type of product the user prefers. Then, the businesses do advertisement showcasing the preferred product to the specific user. Moreover, it allows businesses to make decisions. Using this information, sentiment of the user are understood to make it favorable to their product selling. Social media has also become a common platform to advertise not only products, but also talent, movies, music, messages. New movie reviews and comments of the users can be analyzed to know the reaction of people and the level of excitement which may lead to flop or hit of a movie. Similarly, social media is used to showcase their talent such as dance, singing, mimicry etc. performed by common people. Along with data the user using social media is also increasing with a very high speed. In the near future the data produced by the social media are also going to reach Exascale. The business analytic has to be prepared for the Exascale Big data processing. It should be intelligent enough to differentiate between the useful and not useful data. Another important point is in Exascale era the social media service provider has to protect their users from cyber bullies, cyber crime, etc.  Protecting clients from damaging, unethical and illegal data (messages, music, videos etc.). Analyzing and finding such harmful content along with maintaining the privacy and integrity of the user is going to be a difficult task for the social media service provider. 

The history is going to repeat itself. When the Big data era came the traditional methods are unable to process the Big data. Again, when the Exascale era will come the Big Data technologies will be unable to process the data. In such situation, only storing data will be meaningless. Many business fields are gaining profits by using the knowledge obtained after processing of data. They will suffer huge losses. As explained earlier controlling the data will also be difficult. In social media, when the Big data technologies will not work, then their users data will be left vulnerable. Hence, Exascale Big Data requires a Big computing system before the onset of the Exascale era.

\subsection{Future Application}
Big computing helps in solving many complex problems and simulations. Some fields that have future applications of Big computing have advanced material science, new energy solution, urban planning, cosmology, astrophysics and many more \cite{messina}. Discovery of new material with required characteristics in the field of advanced material science needs solution to many complex calculations. Known compounds are combined to obtain the desired properties. These calculations help in simulating the behavior of the obtained material in the nature. Currently, classical simulation and deep learning are used for the research. And, with Big Computing more complex design and calculation can be performed. The challenge of finding an alternative energy sources can also be solved using Big computing. For example, suppose in a wind farm the speed of wind is very less. Then, for continuous production of electricity quickly the fossil fuel generator is started. Predicting such events manually is impossible. Such actions need heavy and Big computation. Big Computing helps in understanding the chemical composition of the fossil fuel. This helps in optimizing the energy production and reducing pollution. Big Computing also will play an important role in increasing the quality of life. Big computing processes Exascale data to optimize infrastructure access and usage choices. Data is collected from several sources such as sensors, databases, survey done by government. These data scale to Exabytes. Big computing analyze and simulate thousands of scenarios. It obtains new planned city or restructuring existing large cities. Another field of science that desperately need Big computing system is cosmology \cite{cosmological} and astrophysics. This field of science is interested in the evolution of the universe and creation, and destruction of stars. Research in such field heavily depends on complex calculations and simulations. Hence, requires the help of Big Computing to unfold the hidden mysteries of the universe. 

Initially the focus of the scientist was to obtain the whole genome to visualize the genome structure. Recently scientist has realized that apart from storage there is a requirement for processing of such Exascale data. After a genome is assembled, computation is performed to analyze the correlation of mutation with the disease or search for the history of evolution of organisms. But such computation requires pairwise comparison between genomes. This leads to computation to a quadratic scale. Hence, with moving of Exascale Big Data the science world move towards Big Computing capable of Exascale computing. The DNA with its complex network of chemical composition is an example of Big Graph. Such Big Graph with higher dimension requires Big Computing to estimate the parameter and optimize the metabolic model. This metabolic model helps with discovering important genes that encode proteins. It is useful in bio-manufacturing and other industries. Big Computing will also show the path to personalized medicine. The effect of an administered drug will be studied using Big Computing. Big Computing will help with couple drug-induced perturbations of channel behavior and cellular action potential model. Moreover, in case of a drug behaviour in patient’s heart. The MRI images are used to obtain a 3D image of the heart. But such analysis needs the support of Big Computing.

\subsection{Hardware and Software Requirements}
Big computing system at Exascale level requires both upgraded hardware and software \cite{Petcu}. First such systems have to solve many issues such as synchronization, load balancing, failure handling, and distributed memory management. One main requisite of Big computing systems is an efficient design for expressing task dependencies and inter-task parallelism. The system also requires concurrent communication among a very large set of tasks. Moreover, heterogeneous computation components (eg. CPU and GPU both integrated in a single chip) are also proposed. However, a heterogeneous target computing infrastructures arises more issues. Big computing system have to partition the Exascale data to small dataset and perform parallel operations numbered to thousands or millions. Exascale data processing also requires a new set of languages that are capable of extensive parallelism. Many languages are developed based on data-centric methods. Some examples are X10 \cite{x10}, UPC \cite{UPC}, and Legion \cite{legion}. Correspondingly, Big computing requires large memory, hardware supporting large distributed system. MPI is capable to handle one million cores. This makes MPI eligible to be included in Big computing systems. Most important issue the Big computing systems will face is handling failure. In future the number of compute nodes is going to increase. And, failure of these nodes will be very common and within a very small duration, mean time to failure of 2 hours \cite{Bautista}. To maintain reliability in Big computing systems, multiple software and hardware techniques need to be deployed to predict crashes and tries to maintain the stability of the system. P2P-MapReduce \cite{p2pmap} is an adaptive MapReduce framework. It manages Namenode failures, Datanode churn and does decentralized job recovery. It helps to provide a reliable MapReduce middleware in a dynamic large-scale distributed system.

\subsection{Data-intensive computation}
Big computing systems will be more data intensive rather than compute-intensive. Hence, more support is required from the storage systems \cite{freeman}. The storage system has to support intensive metadata operations. Optimized file writing is required. Fault tolerance should also be handled by some new techniques of checkpointing. The storage system has to flush the memory for external storage. And, rolling back of memory for checkpoint on occasion of failure should be rare \cite{ Zhao}. Because, it will lead to increment in I/O cost. Jin et al. \cite{Jin} proposed a multi-tiered dynamic data staging that integrates both DRAM and SSD. Solid state disk (SSD) is a non-volatile memory device which is becoming popular. It is deployed as storage buffers to cache temporally the checkpoint data \cite{Prabhakar} or caching data before writing to disks \cite{Li}. The hybrid data staging is able to store large data volume, which exceeds the capacity of currently used DRAM. It has other merits such as it supports code coupling, data sharing and management, and has adopted an application-aware data placement mechanism. 

It is estimated that the file based data exchange method will become infeasible to implement in the Big computing system \cite{Ma}. The cost of I/O will not permit to achieve the performance target required by the Big computing system. Hence, the hardware has to be efficient to provide a solution to this issue. Another constraint is data locality. The main reasons are high latency and insufficient bandwidth. One possible solution is in-situ data analysis \cite{ovsyannikov}. It means moving the computation where the data resides. Two ways are possible to implement in-situ data analysis. First, unified execution by integration of the data processing and data production. The data analysis component access the produced data by making function calls to their shared memory address space. It is becoming popular because it helps to decrease the I/O operation time in data-intensive systems. In addition, overly decreases the end-to-end workflow runtime. Second, implementing data staging. In data staging the workflow coupling is done by converting some compute nodes into dedicated I/O nodes. Some applications can use the memory of these nodes for storage and exchanging intermediate workflow data. In some framework this mechanism is implemented such as DataStager \cite{Abbasi} and PreDatA \cite{PreDatA}. However, it has many issues. API integrated data staging does recompiling after the application compilation. Data staging need extra compute resources.


\subsection{GPU era}
Compared to CPU, GPU provides more peak performance and bandwidth using less power. A Big computing system with GPU will be smaller compared to non-GPU system. Therefore, GPU will be the building block of the Big computing system \cite{Vuduc}. A GPU based Big computing system will require less compute resources to meet the performance target. Moreover, in case of more I/O bound operation the GPU will also provide efficient memory bandwidth. However, in Big computing system the overall consumption of power has to be constraint. Hence, more power efficient GPU accelerators are essential. In GPU based Big computing system, data communication between the host and the GPU is only through GPU memory. In case of parallel application, GPU requires more data access \cite{Che}. It results in less performance and more power consumption. The power consumption by the GPU is based on many factors. Moreover, designing energy efficient scheduling for the GPU is difficult. Because the scheduler is embedded in the device firmware \cite{Keckler}. Separate study is required to make the GPU more efficient because knowledge used to study CPU cannot be used on GPU due to differences in the architecture and organization \cite{Allen}.    

Currently, the idea of combining both the CPU and GPU is proposed. The best features of both are combined to further increase the overall performance. This is termed as heterogeneous Computing (HC). CPUs are efficient in latency-critical applications and GPUs are efficient in throughput-critical applications. The aim of the HC unit will be load balancing between CPU and GPU to keep both in idle state for very less duration. It is planned to embedded both in the same chip. Many similar processors already exist, such as gem5-gpu \cite{Power} and Ivy Bridge \cite{Damaraju}. However, integrating both also have some issues such as \cite{Mittal}: (a) HC can have either discrete or fused architecture, (b) power consumption (c) data transfer and memory bandwidth overhead (d) pipelining data transfer between CPU and GPU (e) efficient algorithm to include high levels of parallelism (f) partition of workload to keep both for very less duration in idle state. The HC is also going to get affected by the issues they individually pose. In addition, the optimization techniques implemented on CPU and GPU individually will not be applicable to HC.

In Big computing system, GPU is going to play a big role to meet the performance peak. Similarly, the advancement made in making the CPU efficient can be utilized to increase the performance. Both GPU-CPU integration also provides another way in achieving an efficient Big computing system. However, above discussed issues need to be solved before there are fully implemented. Finally, they have to be very optimal in power consumption because overall power consumption has to be less to achieve the Big Computing reality.

\subsection{Energy-aware computation}
Power consumption is the most important issue the Big computing system has to solve. Because the amount of power the Big computing system consume is threatening its feasibility \cite{Ahmad}. The world is facing power shortage issue, hence, providing megawatts of energy to a Big computing system is very impractical. For example, when fully utilized China’s Tianhe-2 uses 17 MegaWatt. Currently, it is estimated that a Big computing system on an average requires 20 megawatts of energy. A Big computing system has many power consumption sources. Some examples are (a) cooling of the heat produced by the Big computing system. Its contribution is very high. (b) data storage systems (c) networking (d) data warehouse having the Big computing system. They are also restricted to some specific location due to security and historical reason.  
Achieving lower power consumption need optimization of every components of the Big computing system. Using power efficient computing components (eg. ARM). Data transfer also requires power. So, increasing the locality of data and reducing compute components to reduce the traffic can be considered. Reducing the speed automatically when less data need to be transferred. Variable data transfer speed should be adopted to reduce energy wherever possible. In different software components (application, scheduler, and runtime) energy-conserving mechanisms should be adopted \cite{panda}. 

ECOSCALE \cite{ecoscale} is a novel energy efficient heterogeneous hierarchical architecture. It has a hybrid architecture having many-core+OpenCL programming environment and a runtime component. It partition the physical system into many compute nodes. These compute nodes are interconnected in a tree-like fashion. ECOSCALE defines a contiguous global address space where nodes are hierarchically interconnected by an MPI protocol. The compute nodes further reduces power consumption and provides resilience by implementing a dual stage Memory Management Unit. This unit maps the reconfigurable accelerators to virtual address space. 

Another power consuming factor is the cooling of the heat produced by the system. In Big computing system using traditional methods (Air conditioner) is not sufficient. Hence, new approaches are proposed to cool the Big computing system. Liquid cooling is another solution which provides higher cooling compared to traditional methods. Some examples are spray cooling \cite{SILK}, microchannel \cite{REZANIA} and jet impingement \cite{HOSAIN}. Currently, a hybrid approach is also proposed which combines cooling liquid and traditional method. Galileo \cite{conficoni} is a cooling approach which combines  traditional method and rack-level RDHXs. Rack-level RDHX is mounting  liquid-to-air rear door heat exchangers (RDHX) at rack level. Chen \textit{et al.} \cite{chen} proposed a novel system that does efficient cooling by reusing waste heat. The system combines a plug-type spray cooling system and absorption chiller. 

\section{Conclusion}
\label{con}
The world with a very high speed is moving towards the Exascale era. The near future of Exascale is a horrible future, which is capable of impeding the world with its avalanche of data. And, currently the world is nowhere near to handle such an era. The first step in Exascale is producing a Big computing machine. Before achieving the first step many preparations are required. Just to reach the first step of supercomputer many issues and challenges need to be overcome. Some issues and challenges are energy requirement, hardware failure, fault tolerance, communication and software complexity. In this paper, we have presented Big Computing which is two-fold, namely, data-intensive and task-intensive computation. Therefore, Exascale data and task computing will pose many research challenges and issues which are already discussed in this paper. Energy-aware computing is next big things to do for research. Also, 5G will revolutionize entire industry towards Big Computing. Big Data 2.0 is yet to conceptualize for very large scale data. Similarly, ExaFLOP need to be evolved to reach ZettaFLOP. Thus, we expect Big Computing will be the next big thing.



\bibliographystyle{icstnum}
\bibliography{mybib.bib}
\end{document}